\documentclass[12pt,showpacs,preprintnumbers,amsmath,amssymb]{revtex4}
\usepackage{color} %example: {\color{red} text}
\usepackage{graphicx}% Include figure files
\begin{document}

\newcommand{\eq}[1]{Eq.~(\ref{#1})}  %example: \eq{eq:NFPE}

\title{Multicritical Behavior of Two Coupled Ising Models in the 
Presence of a Random Field}

\vskip \baselineskip

\author{Octavio D. Rodriguez Salmon}
\thanks{E-mail address:octaviors@gmail.com}
\affiliation{Departamento de F\'{\i}sica \\
Universidade Federal do Amazonas, 3000, Japiim \\
69077-000 \hspace{5mm} Manaus - AM \hspace{5mm} Brazil}

\author{Fernando D. Nobre}
\thanks{Corresponding author: E-mail address: fdnobre@cbpf.br}
\affiliation{Centro Brasileiro de Pesquisas F\'{\i}sicas and \\
National Institute of Science and Technology for Complex Systems \\
Rua Xavier Sigaud 150 \\
22290-180 \hspace{5mm} Rio de Janeiro - RJ \hspace{5mm} Brazil}

\date{\today}

%\newpage
%\vskip \baselineskip

\begin{abstract}
\noindent
A system defined by two coupled Ising models, 
with a bimodal random field acting in one of them, 
is investigated. The interactions among variables
of each Ising system
are infinite-ranged, a limit where mean field 
becomes exact.
This model is studied at zero temperature, as 
well as for finite temperatures, representing
physical situations which are 
appropriate for describing  
real systems, such as plastic crystals.
A very rich critical behavior is found, depending
directly on the particular choices of the temperature, 
couplings, and random-field strengths. 
Phase diagrams exhibiting ordered, partially-ordered, 
and disordered phases are analyzed, showing   
the sequence of transitions through 
all these phases, similarly to what occurs in plastic 
crystals. 
Due to the wide variety of critical phenomena
presented by the model, its usefulness for 
describing critical behavior in other substances
is also expected.  

\vskip \baselineskip
%\vspace{2cm}
%\medskip

\noindent
Keywords: Multicritical Phenomena, Random-Field Ising Model, Plastic Crystals.
\pacs{05.70.Fh, 05.70.Jk, 64.60.-i, 64.60.Kw}

\end{abstract}
\maketitle
\newpage

\section{Introduction}

The study of magnetic models has 
generated considerable progresses in the understanding 
of magnetic materials, 
and lately, it has overcome the frontiers of magnetism,
being considered in many areas of knowledge.  
Certainly, the Ising model represents one of the most 
studied and important models of magnetism and
statistical mechanics~\cite{huang,reichl}, 
and it has been employed also to typify a wide variety of 
physical systems, like lattice gases, binary alloys, and 
proteins (with a particular interest in the problem of protein 
folding). 
Although real magnetic systems should be properly 
described by means of Heisenberg spins (i.e.,
three-dimensional variables), many materials are
characterized by anisotropy fields that make these 
spins prefer given directions in space, explaining 
why simple models, characterized by  
binary variables, became so important 
for the area of magnetism. 
Particularly, models defined in terms of Ising variables
have shown the ability for exhibiting a wide variety 
of multicritical behavior
by introducing randomness, and/or competing 
interactions, has attracted the attention of many
researchers~(see, e.g., Refs.~\cite{aharony,mattis,kaufman,nogueira98,% 
nuno08a,nuno08b,salmon1,morais12}). 

Certainly, the simplicity of Ising variables, 
which are very suitable for both analytical and numerical
studies, has led to proposals of important models outside
the scope of magnetism, particularly in the 
area of complex systems.
These models have been successful for describing 
a wide variety of relevant 
features in such systems, and have raised  
interest in many fields, like
financial markets, optimization problems,  
biological membranes, and social behavior.
In some cases, more than one Ising variable have been used, 
especially by considering a coupling between them, as 
proposed within the framework of choice 
theories~\cite{fernandez}, or in plastic 
crystals~\cite{plastic1,brandreview,folmer}.  
In the former case, each set of Ising variables represents
a group of identical individuals, all of which can make two
independent binary choices. 

%%%%%%%%%%%%%%%%%%%%%%%%%%%%
\begin{figure}[htp]
\begin{center}
\includegraphics[height=5.5cm]{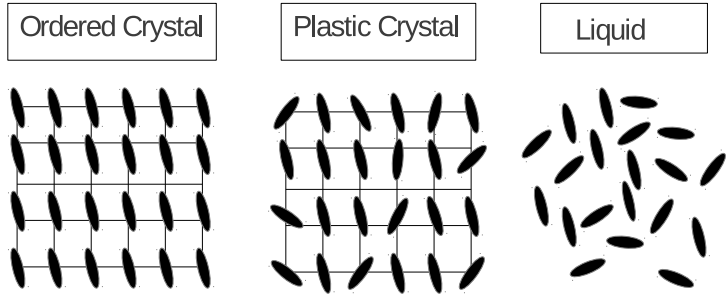}
\end{center}
\vspace{-1cm}
\caption{Illustrative pictures of the three phases as the temperature 
increases, low-temperature (ordered) solid, intermediate 
plastic crystal, and high-temperature (disordered) liquid phase. 
In the plastic state the  centers of mass of the molecules form a 
regular crystalline lattice but the molecules are  
disordered with respect to the orientational degrees of freedom.} 
\label{fig:fasesdecristais}
\end{figure}
%%%%%%%%%%%%%%%%%%%%%%%%%%%%

The so-called plastic 
crystals~\cite{plastic1,brandreview,folmer,michel85,michel87,%
galam87,galam89,salinas1,salinas2} appear as states 
of some compounds considered to be simpler than those of canonical 
glasses, but still presenting rather nontrivial 
relaxation and equilibrium properties. Such a plastic
phase corresponds to an intermediate stable state, between a 
high-temperature (disordered) liquid phase, and a low-temperature
(ordered) solid phase and both transitions, 
namely, liquid-plastic and plastic-solid, are first order. 
In this intermediate phase, the rotational disorder coexists
with a translationally ordered state, characterized by 
the centers of mass of the molecules forming a regular crystalline 
lattice with the molecules presenting disorder in their 
orientational degrees of freedom, as shown 
in Fig.~\ref{fig:fasesdecristais}. 
Many materials undergo a liquid-plastic phase transition, 
where the lower-temperature phase presents such a  
partial orientational order, like the plastic-crystal 
of Fig.~\ref{fig:fasesdecristais}. 
The property of translational invariance makes the plastic crystals
much simpler to be studied from both analytical and numerical 
methods, becoming very useful towards a proper 
understanding of the glass transition~\cite{plastic1,brandreview,folmer}.   
In some plastic-crystal models one introduces a coupling 
between two Ising models, associating each of these 
systems respectively, to the translational and rotational degrees of 
freedom~\cite{galam87,galam89,salinas1,salinas2}, 
as a proposal for explaining satisfactorily 
thermodynamic properties of the plastic phase.  

Accordingly, spin variables $\{t_{i}\}$ and $\{r_{i}\}$ are introduced in 
such a way to mimic translational 
and rotational degrees of freedom of each molecule $i$, respectively. 
The following Hamiltonian is 
considered~\cite{galam87,galam89,salinas1,salinas2}, 

\begin{equation}
\label{eq:hamplastcrystals}
{\cal H} = - J_{t}\sum_{\langle ij \rangle}t_{i}t_{j}
- J_{r} \sum_{\langle ij \rangle}r_{i}r_{j} 
- \sum_{i} (\alpha t_{i} + h_{i})r_{i}~,
\end{equation}

\vskip \baselineskip
\noindent
where $\sum_{\langle ij \rangle}$ represents a sum over 
distinct pairs of nearest-neighbor spins. 
In the first summation, the Ising variables $t_{i}=\pm 1$ 
may characterize two lattices A and B (or occupied and vacant sites). 
One notices that the rotational 
variables $r_{i}$ could be, in principle, continuous variables, 
although the fact that the minimization of the coupling contribution 
$\alpha t_{i}r_{i}$ is achieved 
for $t_{i}r_{i} =1$ ($\alpha>0$), or 
for $t_{i}r_{i} =-1$ ($\alpha<0$), suggests the simpler choice
of binary variables ($r_{i}=\pm 1$) to be appropriate, 
based on the energy minimization requirement.   

In the present model the variables $t_{i}$ and 
$r_{i}$ represent very different characteristics of a 
molecule. Particularly, the rotational variables $r_{i}$
are expected to change more freely than the translational ones; 
for this reason, one introduces a random field acting only 
on the rotational degrees of freedom.  
In fact, the whole contribution $\sum_{i} (\alpha t_{i} + h_{i})r_{i}$
is known to play a fundamental role for the plastic phase of 
ionic plastic crystals, like the alkalicyanides KCN, NaCN and RbCN. 
In spite of its simplicity, the above Hamiltonian is able to capture 
the most relevant features of the plastic-crystal phase, as well 
as the associated phase transitions,
namely, liquid-plastic and plastic-solid ones~\cite{michel85,michel87,%
galam87,galam89,salinas1,salinas2,vives}. 

A system described by a Hamiltonian slightly different 
from the one of~\eq{eq:hamplastcrystals}, in which the 
whole contribution
$\sum_{i} (\alpha t_{i} + h_{i})r_{i}$ was replaced by 
$\sum_{i} \alpha_{i} t_{i}r_{i}$, i.e., with no random 
field acting on variable $r_{i}$ separately, was considered
in Ref.~\cite{salinas2}. In such a work one finds a detailed
analysis of the phase diagrams and order-parameter behavior 
of the corresponding model. However, to our knowledge,   
previous investigations on the model defined 
by~\eq{eq:hamplastcrystals} have not 
considered thoroughly the effects of the random 
field $h_{i}$, with a particular attention to the phase diagrams 
for the case of a randomly distributed bimodal
one, $h_{i}=\pm h_{0}$;   
this represents the main motivation
of the present work. 
In the next section we define the model, determine its free-energy 
density, and describe the 
numerical procedure to be used. 
In Section III we exhibit typical phase diagrams 
and analyze the behavior of the corresponding order parameters,
for both zero and finite temperatures; the ability of the model 
to exhibit a rich variety of phase diagrams, characterized 
by multicritical behavior, is shown. 
Finally, in Section IV we present our main conclusions.  
 
\section{The Model and Free-Energy Density}   

Based on the discussion of the previous section, herein
we consider a system composed by two interacting Ising models, 
described by the Hamiltonian

\begin{equation}
\label{eq:hamiltonian1}
{\cal H}(\{h_{i}\}) = - J_{\sigma} \sum_{(ij)}\sigma_{i}\sigma_{j}
- J_{\tau} \sum_{(ij)}\tau_{i}\tau_{j} + D\sum_{i=1}^{N}\tau_{i}\sigma_{i}
-\sum_{i=1}^{N}h_{i}\tau_{i}~, 
\end{equation}

\vskip \baselineskip
\noindent
where $\sum_{(ij)}$ represent sums over all distinct pairs of spins, 
a limit for which the mean-field approximation becomes exact. Moreover, 
$\tau_{i}= \pm 1$ and $\sigma_{i}= \pm 1$ ($i=1,2, \cdots , N$) depict 
Ising variables,  
$D$ stands for a real parameter, whereas both $J_{\sigma}$ and  
$J_{\tau}$ are positive coupling constants, which will be 
restricted herein to 
the symmetric case, $J_{\sigma}=J_{\tau}=J>0$. Although this later 
condition may seem as a rather artificial simplification of the 
Hamiltonian in~\eq{eq:hamplastcrystals}, the application of a 
random field $h_{i}$ acting separately on one set of variables, will 
produce the expected distinct physical behavior associated with  
$\{ \tau_{i} \}$ and $\{ \sigma_{i} \}$. The random fields 
$\{ h_{i} \}$ will be considered as following 
a symmetric bimodal probability distribution function, 

\begin{equation}
\label{eq:hpdf}
 P(h_{i}) = \frac{1}{2} \, \delta(h_{i}-h_{0}) +\frac{1}{2} \, \delta(h_{i}+h_{0})~.  
\end{equation}
 
\vskip \baselineskip
\noindent
The infinite-range character of the interactions allows one to write the above 
Hamiltonian in the form

\begin{equation}
\label{eq:hamiltonian2}
{\cal H}(\{h_{i}\})= - \frac{J}{2N}{\left (\sum_{i=1}^{N}\sigma_{i} \right )}^{2} 
- \frac{J}{2N}{\left (\sum_{i=1}^{N}\tau_{i} \right )}^{2} 
+D\sum_{i=1}^{N}\tau_{i}\sigma_{i} -\sum_{i=1}^{N}h_{i}\tau_{i}~,
\end{equation} 

\vskip \baselineskip
\noindent
from which one may calculate the partition function associated with 
a particular configuration of the fields $\{ h_{i}\}$, 

\begin{equation}
Z(\{h_{i}\}) =  {\rm Tr} \exp \left[- \beta {\cal H}(\{h_{i}\})  \right]~, 
\end{equation}

\vskip \baselineskip
\noindent
where $\beta=1/(kT)$ and 
${\rm Tr} \equiv {\rm Tr}_{\{ \tau_{i},\sigma_{i}=\pm 1 \}} $ indicates a sum over 
all spin configurations. One can now make use of 
the Hubbbard-Stratonovich transformation~\cite{dotsenkobook,nishimoribook}
to linearize the quadratic terms, 

\begin{equation}
Z(\{h_{i}\}) =  \frac{1}{\pi} \int_{-\infty}^{\infty}dx dy \exp(-x^{2}-y^{2}) 
\prod_{i=1}^{N} {\rm Tr} \exp [ H_{i}(\tau,\sigma)]~,
\end{equation}
 
\vskip \baselineskip
\noindent
where $H_{i}(\tau,\sigma)$ depends on the random 
fields $\{ h_{i}\}$, 
as well as on the spin variables, being given by

\begin{equation}
H_{i}(\tau,\sigma) = \sqrt{\frac{2\beta J}{N}} \ x \tau + \sqrt{\frac{2\beta J}{N}} \ y \sigma 
- \beta D \tau \sigma + \beta h_{i} \tau~.   
\end{equation}

\vskip \baselineskip
\noindent
Performing the trace over the spins and defining new variables, related to 
the respective order parameters, 

\begin{equation}
\label{eq:mtausigma}
m_{\tau} = \sqrt{\frac{2kT}{JN}} \ x~; \qquad 
m_{\sigma} = \sqrt{\frac{2kT}{JN}} \ y~, 
\end{equation}

\vskip \baselineskip
\noindent
one obtains

\begin{equation}
Z(\{h_{i}\})= \frac{\beta J N}{2 \pi} \int_{-\infty}^{\infty} dm_{\tau} dm_{\sigma} \exp[N g_{i} (m_{\tau},m_{\sigma})]~,   
\end{equation}

\vskip \baselineskip
\noindent
where

\begin{eqnarray}
g_{i}(m_{\tau},m_{\sigma}) &=& - \frac{1}{2} \beta J m_{\tau}^{2} 
- \frac{1}{2} \beta J m_{\sigma}^{2} + \log \left \{ 
2e^{-\beta D} \cosh[\beta J(m_{\tau}+m_{\sigma}+h_{i}/J)]
\right. \nonumber \\ \nonumber \\
\label{eq:gimtausigma}
&+& \left. 2e^{\beta D} \cosh[\beta J(m_{\tau}-m_{\sigma}+h_{i}/J)] \right \}~.
\end{eqnarray}

\vskip \baselineskip
\noindent
Now, one takes the thermodynamic limit ($N \rightarrow \infty$), and uses the saddle-point
method to obtain 

\begin{equation}
Z = \displaystyle \frac{\beta J N}{2 \pi} \int_{-\infty}^{\infty} dm_{\tau} dm_{\sigma} 
\exp[-N \beta f(m_{\tau},m_{\sigma})]~, 
\end{equation}

\vskip \baselineskip
\noindent
where the free-energy density functional $f(m_{\tau},m_{\sigma})$ results 
from a quenched average of 
$g_{i}(m_{\tau},m_{\sigma})$ in~\eq{eq:gimtausigma}, over the 
bimodal probability distribution of~\eq{eq:hpdf}, 

\begin{equation}
\label{eq:freeenergy}
f(m_{\tau},m_{\sigma}) = \displaystyle  \frac{1}{2} J m_{\tau}^{2} 
+ \frac{1}{2}  J m_{\sigma}^{2} - \frac{1}{2\beta}\log Q(h_{0})  
- \frac{1}{2\beta}\log Q(-h_{0})~, 
\end{equation}

\vskip \baselineskip
\noindent
with

\begin{equation}
Q(h_{0}) = 2e^{-\beta D} \cosh[\beta J(m_{\tau}+m_{\sigma} + h_{0}/J)]
+2e^{\beta D} \cosh[\beta J(m_{\tau}-m_{\sigma} + h_{0}/J)]~. 
\end{equation}

\vskip \baselineskip
\noindent
The extremization of the free-energy density above with respect to the 
parameters $m_{\tau}$ and $m_{\sigma}$ yields the following equations of state, 

\begin{eqnarray}
\label{eq:mtau}
m_{\tau} &=& \frac{1}{2} \frac{R_{+}(h_{0})}{Q(h_{0})} 
+ \frac{1}{2} \frac{R_{+}(-h_{0})}{Q(-h_{0})}~, 
\\ \nonumber \\
\label{eq:msigma}
m_{\sigma} &=& \frac{1}{2} \frac{R_{-}(h_{0})}{Q(h_{0})} 
+ \frac{1}{2} \frac{R_{-}(-h_{0})}{Q(-h_{0})}~,
\end{eqnarray}

\vskip \baselineskip
\noindent
where 

\begin{equation}
R_{\pm}(h_{0}) = e^{-\beta D} \sinh[\beta J(m_{\tau}+m_{\sigma} + h_{0}/J)] 
\pm e^{\beta D} \sinh[\beta J(m_{\tau}-m_{\sigma} +h_{0}/J)]~. 
\end{equation}

\vskip \baselineskip
\noindent

In the following section we present numerical results for the 
order parameters and phase diagrams of the model, at both
zero and finite temperatures. 
All phase diagrams are represented 
by rescaling conveniently the energy parameters of the system, namely,     
$kT/J$, $h_{0}/J$ and $D/J$. 
Therefore, for given values of these dimensionless parameters,
the equations of state [Eqs.(\ref{eq:mtau}) and~(\ref{eq:msigma})] 
are solved numerically for $m_{\tau}$ and $m_{\sigma}$. 
In order to avoid metastable states, all solutions obtained for 
$m_{\tau} \in [-1,1]$ and $m_{\sigma} \in [-1,1]$ are 
substituted in~\eq{eq:freeenergy},
to check for the minimization of the free-energy density. 
The continuous (second order) critical frontiers are found by the set 
of input values for which the order parameters fall continuously down to 
zero, whereas the first-order frontiers were found through 
Maxwell constructions. 

Both ordered ($m_{\tau} \neq 0$ and $m_{\sigma} \neq 0$) 
and partially-ordered ($m_{\tau}=0$ and $m_{\sigma} \neq 0$) 
phases have appeared in our analysis, and will be labeled 
accordingly.  
The usual paramagnetic phase ({\bf P}),  
given by $m_{\tau}=m_{\sigma}=0$, always occurs for sufficiently
high temperatures. 
A wide variety of critical points appeared in our analysis 
(herein we follow the classification due to Griffiths~\cite{griffiths}): 
(i) a tricritical point signals the encounter of a continuous frontier 
with a first-order line with no change of slope; 
(ii) an ordered critical point corresponds to an isolated critical
point inside the ordered region, terminating a first-order line that
separates two distinct ordered phases;
(ii) a triple point, where three distinct phases coexist, signaling the 
encounter of three first-order critical frontiers. 
In the phase diagrams we shall use distinct symbols and 
representations for the critical points and frontiers, as described below.   

\begin{itemize}

\item Continuous (second order) critical frontier: continuous line;

\item First-order critical frontier: dotted line;

\item Tricritical point: located by a black circle;

\item Ordered critical point: located by a black asterisk;

\item Triple point: located by an empty triangle.

\end{itemize}

\section{Phase Diagrams and Behavior of Order Parameters}

\subsection{Zero-Temperature Analisis}

\begin{figure}[htp]
\begin{center}
\includegraphics[height=5.5cm]{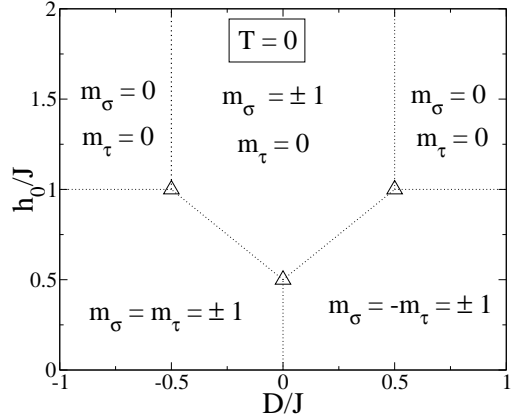}
\end{center}
%\vspace{-1cm}
\caption{Phase diagram of the model defined by Hamiltonian 
of~\eq{eq:hamiltonian2}, at zero temperature. All critical frontiers 
represent first-order phase transitions; the empty triangles denote 
triple points.}
\label{fig:groundstate}
\end{figure}

At $T=0$, one has to analyze the different spin orderings that 
minimize the Hamiltonian of~\eq{eq:hamiltonian2}.
Due to the coupling between the two 
sets of spins, the minimum-energy configurations will correspond to  
$\{\tau_{i}\}$ and $\{\sigma_{i}\}$ antiparallel ($D>0$), or parallel ($D<0$).
Therefore, in the absence of random fields ($h_{0}=0$) one should have 
$m_{\tau}=-m_{\sigma}$ ($D>0$), and 
$m_{\tau}=m_{\sigma}$ ($D<0$),  where $m_{\sigma}=\pm1$.  
However, when random fields act on the $\{\tau_{i}\}$ spins, there will 
be a competition between these fields and the coupling parameter $D$,
leading to several phases, as represented in Fig.~\ref{fig:groundstate}, 
in the plane $h_{0}/J$ versus $D/J$. One finds three ordered 
phases for sufficiently low values of $h_{0}/J$
and $|D|/J$, in addition to {\bf P} phases for $(|D|/J)>0.5$ and $(h_{0}/J)>1$. 
All frontiers shown in Fig.~\ref{fig:groundstate} are first-order critical lines. 

%%%%%%%%%%%%%%%%%%%%%%%%%%%%
%%%%%%%%%%%%%%%%%%%%%%%%%%%%

When $(h_{0}/J) \leq 1/2$ one finds ordered phases for all values of $D/J$, 
with a vertical straight line at $D=0$ separating the 
symmetric state ($D<0$), where $m_{\tau}=m_{\sigma}$, from the   
antisymmetric one ($D>0$), characterized by $m_{\tau}=-m_{\sigma}$. 
Two critical frontiers (symmetric under a reflection operation)
emerge from the triple point at 
$(D/J)=0.0$ and $(h_{0}/J)=0.5$, given, respectively, by 
$(h_{0}/J)=0.5 + (D/J)$ for $D>0$, and 
$(h_{0}/J)=0.5 - (D/J)$ for $D<0$.
These critical frontiers terminate at $(h_{0}/J)=1.0$ and
separate the low random-field-valued ordered phases from 
a partially-ordered 
phase, given by $m_{\tau}=0$ and $m_{\sigma}= \pm 1$.   
As shown in Fig.~\ref{fig:groundstate}, three triple points
appear, each of them signaling the encounter 
of three first-order lines, characterized by a coexistence of three phases, 
defined by distinct values of the magnetizations $m_{\tau}$ and
$m_{\sigma}$, as described below. 

\begin{itemize}

\item $[(D/J)=-0.5$ and $(h_{0}/J)=1.0]$~:  
$(m_{\tau},m_{\sigma})=\{ (0,0);(0,\pm 1); (\pm 1, \pm 1) \}$. 

\item $[(D/J)=0.5$ and $(h_{0}/J)=1.0]$~:
$(m_{\tau},m_{\sigma})=\{ (0,0);(0,\pm 1); (\pm 1, \mp 1) \}$. 

\item  $[(D/J)=0.0$ and $(h_{0}/J)=0.5]$~:
$(m_{\tau},m_{\sigma})=\{ (\pm 1,\pm 1);(\pm 1, \mp 1); (0, \pm 1) \}$. 

\end{itemize}

Such a rich critical behavior shown for $T=0$ suggests that interesting 
phase diagrams should occur when the temperature is taken 
into account. From now on, we investigate the model 
defined by the Hamiltonian of~\eq{eq:hamiltonian2} for finite temperatures. 

\subsection{Finite-Temperature Analysis}

As shown above, the zero-temperature phase diagram presents
a reflection symmetry with respect 
to $D=0$ (cf. Fig.~\ref{fig:groundstate}). 
The only difference between the two sides of this phase
diagram concerns the magnetization solutions  
characterizing the ordered phases for low random-field values,
where one has  
$m_{\tau}=-m_{\sigma}$ ($D>0$), or 
$m_{\tau}=m_{\sigma}$ ($D<0$).  
These results come as a consequence
of the symmetry of the Hamiltonian of~\eq{eq:hamiltonian2}, 
which remains unchanged under the operations, 
$D \rightarrow -D$, $\sigma_{i} \rightarrow -\sigma_{i}$ $(\forall i)$, or 
$D \rightarrow -D$, $\tau_{i} \rightarrow -\tau_{i}$,
$h_{i} \rightarrow -h_{i}$ $(\forall i)$. 
Hence, the finite-temperature phase diagrams should present similar 
symmetries with respect to a change $D \rightarrow -D$. From now on, 
for the sake of simplicity, we will restrict ourselves to the 
case $(D/J) \geq 0$, for which the zero-temperature and 
low-random-field magnetizations present 
opposite signals, as shown in Fig.~\ref{fig:groundstate}, i.e., 
$m_{\tau}=-m_{\sigma}$~. 

%%%%%%%%%%%%%%%%%%%%%%%%%%%%
\begin{figure}[htp]
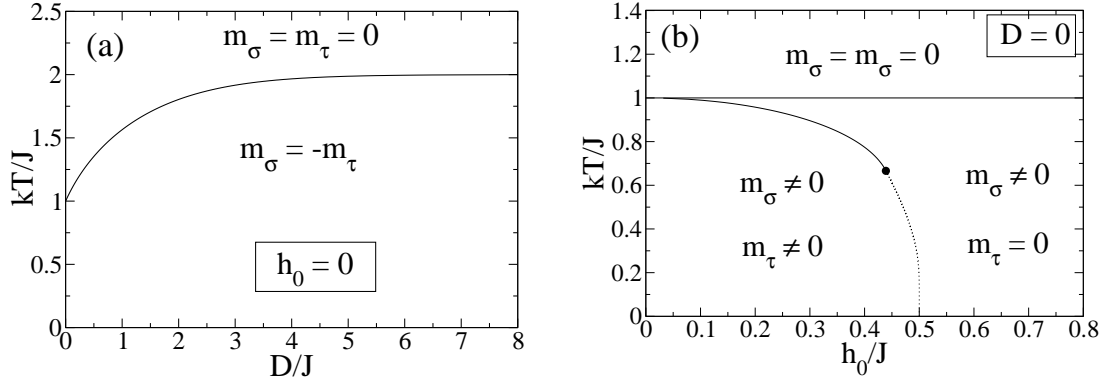

\begin{center}
\vspace{.5cm}
%\includegraphics[height=5.5cm]{figures/figure3a.eps}
%\hspace{0.5cm} 
%\includegraphics[height=5.5cm]{figures/figure3b.eps}
%\includegraphics[height=5.5cm]{figures/figure3a.pdf}
%\hspace{0.5cm} 
%\includegraphics[height=5.5cm]{figures/figure3b.pdf}
\includegraphics[height=5cm]{figure3a.eps}
\hspace{0.5cm} 
\includegraphics[height=5cm]{figure3b.eps}
\end{center}
\vspace{-.5cm}
\caption{Phase diagrams of the model
defined by the Hamiltonian of~\eq{eq:hamiltonian2} in two 
particular cases:  
(a) The plane of dimensionless variables $kT/J$ versus $D/J$, 
in the absence of random fields $(h_{0}=0)$; 
(b) The plane of dimensionless variables $kT/J$ 
versus $h_{0}/J$, for $D=0$. 
The full lines represent continuous phase transitions,
whereas the dotted line stands for a 
first-order critical frontier.
For sufficiently high temperatures one finds a 
paramagnetic phase ({\bf P}), whereas  
the magnetizations $m_{\tau}$ and 
$m_{\sigma}$ become nonzero 
by lowering the temperature.  
In case (b), two low-temperature phases appear, namely,
the ordered (lower values of $h_{0}$) and 
the partially-ordered (higher values of $h_{0}$).
These two phases are separated by a continuous
critical frontier (higher temperatures), which turns into 
a first-order critical line (lower temperatures) at a tricritical
point (black circle). The type of phase 
diagram exhibited in (b) will be called herein of topology I.}
\label{fig:tdh00}
\end{figure}
%%%%%%%%%%%%%%%%%%%%%%%%%%%%

In Fig.~\ref{fig:tdh00} we exhibit phase diagrams of the model
in two particular cases, namely, in the absence of fields $(h_{0}=0)$
[Fig.~\ref{fig:tdh00}(a)] and for zero coupling 
$(D=0)$ [Fig.~\ref{fig:tdh00}(b)]. 
These figures provide useful 
reference data in the numerical procedure 
to be employed for constructing phase diagrams in
more general situations, e.g., in the plane 
$kT/J$ versus $h_{0}/J$, for several values of $(D/J)>0$. 

In Fig.~\ref{fig:tdh00}(a) we present the phase diagram 
of the model in the plane of dimensionless variables $kT/J$ versus $D/J$, 
in the absence of random fields ($h_{0}=0$), 
where one sees the point $D=0$ that corresponds 
to two noninteracting Ising models, leading to the well-known mean-field
critical temperature of the Ising model [$(kT_{c}/J)=1$]. 
Also in Fig.~\ref{fig:tdh00}(a), 
the ordered solution $m_{\tau}=-m_{\sigma}$ 
minimizes the free energy at low temperatures
for any $D>0$; a second-order frontier separates this ordered phase 
from the paramagnetic one that appears for sufficiently high temperatures. 
For high values of $D/J$ one sees that this critical frontier approaches 
asymptotically $(kT/J) = 2$. 
Since the application of a random field results in 
a decrease of the critical temperature, when compared with the one
of the case $h_{0}=0$~\cite{aharony,mattis,kaufman}, 
the result of Fig.~\ref{fig:tdh00}(a) shows that no 
ordered phase should occur for $h_{0}>0$ and $(kT/J)>2$. 

The phase diagram for $D=0$ is shown in 
the plane of dimensionless variables $kT/J$ 
versus $h_{0}/J$ in Fig.~\ref{fig:tdh00}(b). 
The {\bf P} phase occurs for $(kT/J)>1$, whereas 
for $(kT/J)<1$ two phases appear, namely, 
the ordered one (characterized by  
$m_{\sigma} \neq 0$ and $m_{\tau} \neq 0$, with 
$|m_{\sigma}| \geq |m_{\tau}|$), 
as well as the partially-ordered phase 
($m_{\sigma} \neq 0$ and $m_{\tau} = 0$). 
Since the two Ising models are uncorrelated for $D=0$
and the random fields act only on the $\{\tau_{i}\}$ 
variables, one finds that the critical behavior associated 
with variables $\{\sigma_{i}\}$ and $\{\tau_{i}\}$ occur 
independently: 
(i) The variables $\{\sigma_{i}\}$ order at $(kT/J)=1$, for 
all values of $h_{0}$;  
(ii) The critical frontier shown in
Fig.~\ref{fig:tdh00}(b), separating the two
low-temperature phases, is characteristic of an 
Ising ferromagnet in the presence of a bimodal 
random field~\cite{aharony}. The black circle
denotes a tricritical point, where the higher-temperature
continuous frontier meets the lower-temperature
first-order critical line. The type of phase 
diagram exhibited in Fig.~\ref{fig:tdh00}(b)
will be referred herein as topology I.  

%%%%%%%%%%%%%%%%%%%%%%%%%%%%
\begin{figure}[htp]
\begin{center}
%\includegraphics[height=5.5cm]{figures/figure4a.eps}
%\hspace{0.5cm} 
%\includegraphics[height=5.5cm]{figures/figure4b.eps}
%\includegraphics[height=5.5cm]{figures/figure4c.eps}
%\hspace{0.5cm}
%\includegraphics[height=5.5cm]{figures/figure4d.eps}
%\includegraphics[height=5.5cm]{figures/figure4a.pdf}
%\hspace{0.1cm} 
%\includegraphics[height=5.4cm]{figures/figure4b.pdf} \\
%\vspace{0.5cm}
%\includegraphics[height=5.4cm]{figures/figure4c.pdf}
%\hspace{0.1cm}
%\includegraphics[height=5.4cm]{figures/figure4d.pdf}
\includegraphics[height=7.0cm,clip,angle=-90]{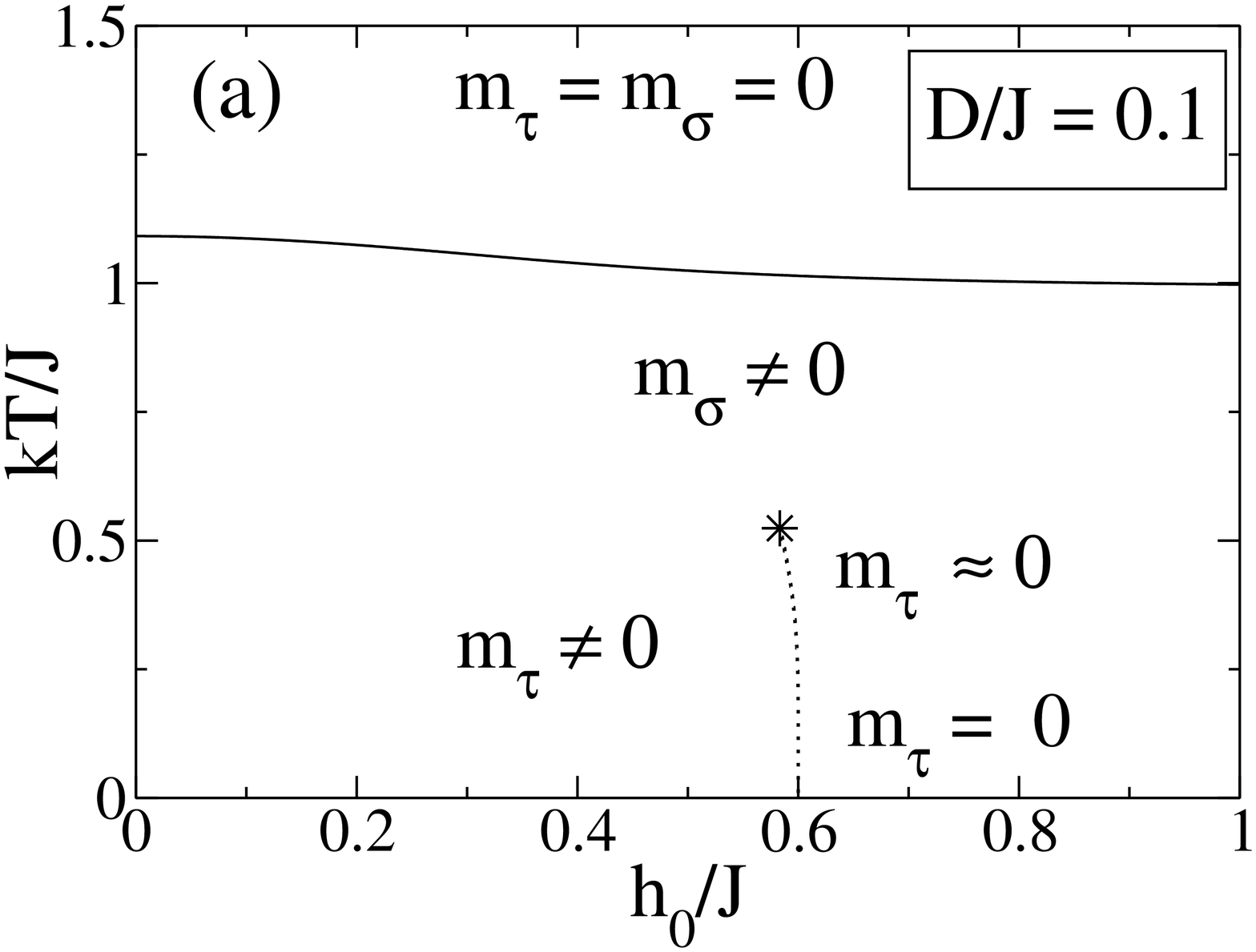}
\hspace{0.1cm} 
\includegraphics[height=7.0cm,clip,angle=-90]{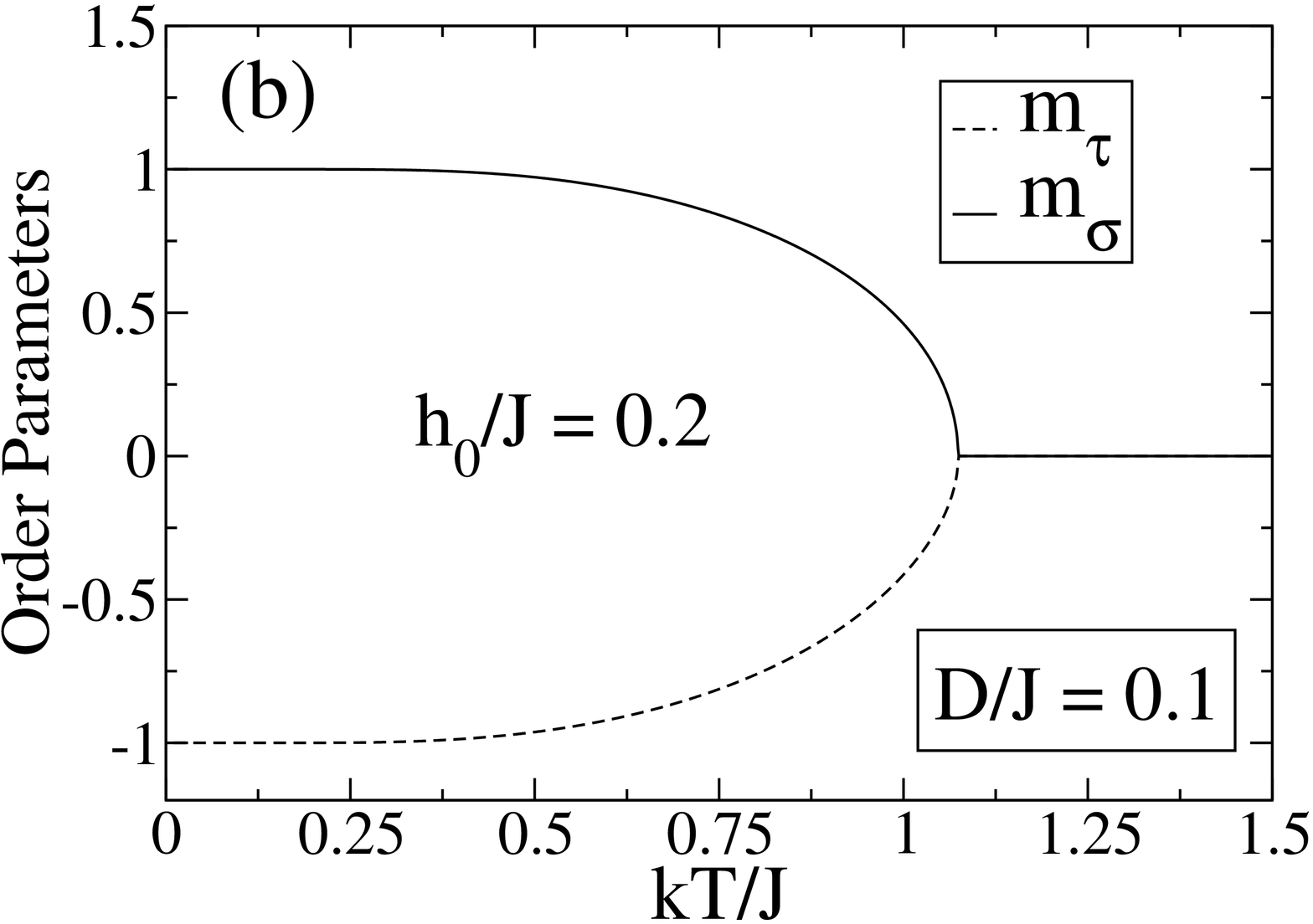} \\
\vspace{0.5cm} \hspace{-0.5cm}
\includegraphics[height=4.5cm,clip]{figure4c.eps}
\hspace{1.0cm}
\includegraphics[height=4.5cm,clip]{figure4d.eps}
\end{center}
\vspace{-.2cm}
\caption{Phase diagram and order parameters in the case
$(D/J)=0.1$. 
(a) Phase diagram in the plane of dimensionless variables $kT/J$ 
versus $h_{0}/J$. At low temperatures, a first-order
critical frontier that terminates in an 
ordered critical point (black asterisk) separates 
the ordered phase (lower values of $h_{0}/J$) from 
the partially-ordered phase (higher values of $h_{0}/J$); 
this type of phase 
diagram will be referred herein as topology II. 
The order parameters $m_{\tau}$ and $m_{\sigma}$
are represented versus the dimensionless temperature 
$kT/J$ for typical values of $h_{0}/J$: 
(b) As one goes through the ordered 
phase (low temperatures) to the {\bf P} phase;  
(c) As one goes through the first-order critical 
frontier, which separates the two ordered phases, 
up to the {\bf P} phase;  
(d) As one goes through the partially-ordered phase 
(slightly to the right of the first-order critical frontier) up 
to the {\bf P} phase. Equivalent solutions exist by 
inverting the signs of $m_{\tau}$ and $m_{\sigma}$.}
\label{fig:d01}
\end{figure}
%%%%%%%%%%%%%%%%%%%%%%%%%%%%

The effects of a small interaction [$(D/J)=0.1$] 
between the variables $\{\sigma_{i}\}$ and 
$\{\tau_{i}\}$ are presented in Fig.~\ref{fig:d01}, where
one sees that the topology I [Fig.~\ref{fig:tdh00}(b)] goes 
through substantial changes, as shown 
in Fig.~\ref{fig:d01}(a) (to be called herein as topology II).
As expected from the behavior presented 
in Fig.~\ref{fig:tdh00}(a), one notices that  
the border of the {\bf P} phase (a continuous frontier) 
is shifted to higher temperatures. 
However, the most significant difference between
topologies I and II consists in 
the low-temperature frontier
separating the ordered and partially-ordered phases. 
Particularly, the continuous frontier, as well
as the tricritical point shown 
in Fig.~\ref{fig:tdh00}(b), give place to an 
ordered critical point~\cite{griffiths}, at which
the low-temperature first-order critical 
frontier terminates. 
Such a topology has been found also in some 
random magnetic systems, like the Ising and Blume-Capel
models, subject to random fields and/or 
dilution~\cite{kaufman,salmon1,salmon2,benyoussef,
carneiro,kaufmankanner}.  
In the present model, we verified that topology II holds 
for any $0<(D/J)<1/2$, with 
the first-order frontier starting at zero temperature and 
$(h_{0}/J)=(D/J)+1/2$, which in Fig.~\ref{fig:d01}(a)
corresponds to $(h_{0}/J)=0.6$. Such a first-order line
essentially affects the parameter $m_{\tau}$, as will be 
discussed next. 

In Figs.~\ref{fig:d01}(b)--(d) the order parameters 
$m_{\tau}$ and $m_{\sigma}$ are exhibited versus   
$kT/J$ for conveniently chosen values of $h_{0}/J$, 
corresponding to distinct physical situations of the 
phase diagram for $(D/J)=0.1$. 
A curious behavior is 
presented by the magnetization 
$m_{\tau}$ by varying $h_{0}/J$, and more 
particularly, around the first-order critical line. 
For $(h_{0}/J)=0.59$ [Fig.~\ref{fig:d01}(c)], 
one starts at low temperatures
essentially to the left of the critical frontier and by increasing 
$kT/J$ one crosses this critical frontier at $(kT/J)=0.499$, 
very close to the ordered critical point. 
At this crossing point, 
$|m_{\tau}|$ presents an abrupt decrease, i.e., 
a discontinuity, corresponding 
to a change to the partially-ordered phase; on 
the other hand, the magnetization $m_{\sigma}$
remains unaffected when going through this critical frontier.  
For higher temperatures, 
$|m_{\tau}|$ becomes very small, but still finite,
turning up zero only at the {\bf P} boundary; in fact, 
the whole region around the ordered critical point
is characterized by a finite small value of $|m_{\tau}|$. 
Another unusual effect is presented in 
Fig.~\ref{fig:d01}(d), for which $(h_{0}/J)=0.65$, i.e., 
slightly to the right of the first-order critical frontier: 
the order parameter $m_{\tau}$ is zero
for low temperatures, but becomes nonzero by increasing the 
temperature, as one becomes closer to the critical ordered 
point. This rather curious phenomenon is directly related to  
the correlation between the variables $\{\sigma_{i}\}$ and 
$\{\tau_{i}\}$: since for $(kT/J) \approx 0.5$ the magnetization
$m_{\sigma}$ is still very close to its maximum value, 
a small value for $|m_{\tau}|$ is induced, so that both 
order parameters go to zero together only at the {\bf P} frontier.  

Behind the results presented in Figs.~\ref{fig:d01}(a)--(d) 
one finds a very interesting feature, namely, the 
possibility of going continuously from the ordered phase to the 
partially-ordered phase by circumventing the ordered critical point.
This is analogous to what happens in many substances, e.g., water,
where one goes continuously (with no latent heat) 
from the liquid to the gas 
phase by circumventing a critical end point~\cite{huang,reichl}.       

%%%%%%%%%%%%%%%%%%%%%%%%%%%%
\begin{figure}[htp]
\begin{center}
%\includegraphics[width=0.45\textwidth,angle=0]{figures/figure5a.eps}
%\hspace{0.2cm}
%\includegraphics[width=0.45\textwidth,angle=0]{figures/figure5b.eps}
%\includegraphics[width=0.45\textwidth,angle=0]{figures/figure5a.pdf}
%\hspace{0.2cm}
%\includegraphics[width=0.45\textwidth,angle=0]{figures/figure5b.pdf}
%\includegraphics[height=5.5cm]{figures/figure5a.pdf}
%\hspace{0.2cm}
%\includegraphics[height=5.5cm]{figures/figure5b.pdf}
\includegraphics[height=6.5cm,angle=-90]{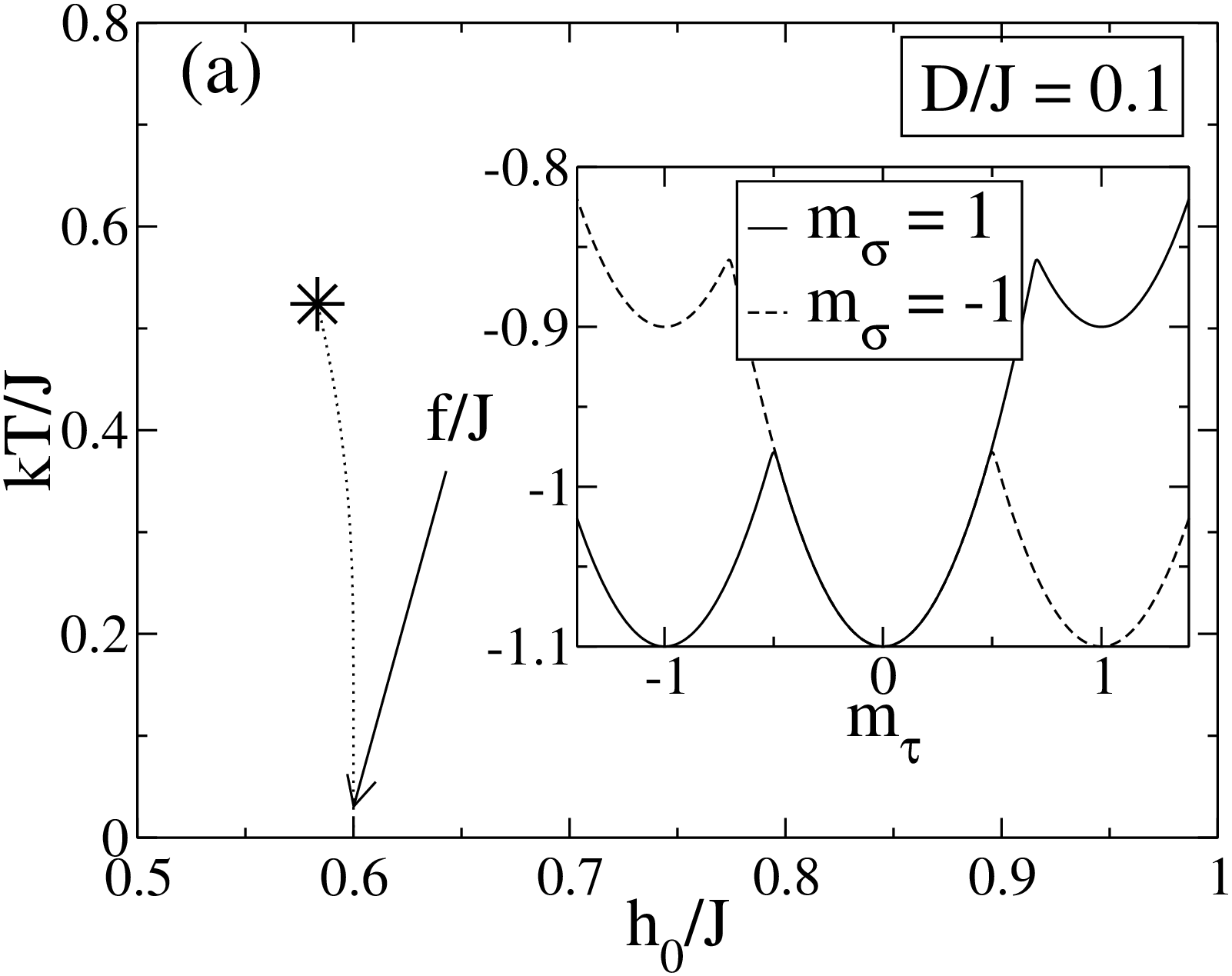}
\hspace{0.2cm}
\includegraphics[height=6.5cm,angle=-90]{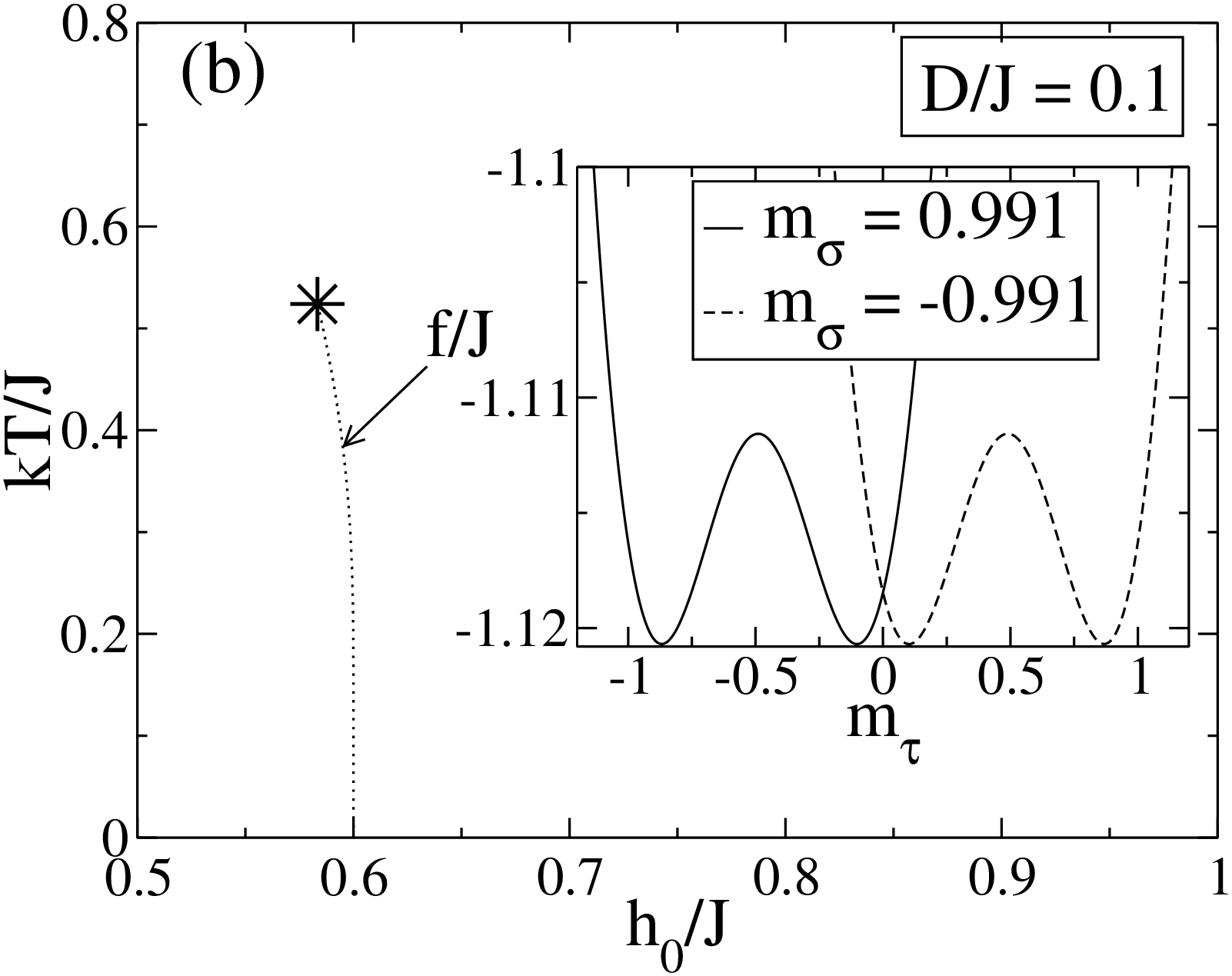}
\end{center}
\vspace{-.5cm}
\caption{The first-order critical line in Fig.~\ref{fig:d01}(a), 
corresponding to $(D/J)=0.1$, is amplified, and 
the dimensionless free-energy density $f/J$ of~\eq{eq:freeenergy} 
(shown in the insets) is analyzed
at two distinct points along this frontier:
(a) A low-temperature point located at $[(h_{0}/J)=0.599,(kT/J)=0.010]$, showing the 
coexistence of the ordered ($|m_{\tau}|=1$) and partially-ordered ($m_{\tau}=0$)
solutions;
(b) A higher-temperature point located at $[(h_{0}/J)=0.594,(kT/J)=0.387]$, 
showing the coexistence of solutions with $|m_{\tau}|>0$, namely, 
$|m_{\tau}|=0.868$ and $|m_{\tau}|=0.1$. 
In both cases (a) and (b) the free energy presents four minima,
associated with distinct pairs of solutions
$(m_{\tau},m_{\sigma})$: the full lines show the two minima 
with positive $m_{\sigma}$, whereas the dashed lines correspond
to the two minima with negative $m_{\sigma}$.} 
\label{fig:freeenergyd01}
\end{figure}
%%%%%%%%%%%%%%%%%%%%%%%%%%%%

In Fig.~\ref{fig:freeenergyd01} the free-energy density of~\eq{eq:freeenergy}
is analyzed at two different points along the first-order critical frontier of 
Fig.~\ref{fig:d01}(a), namely, a low-temperature 
one [Fig.~\ref{fig:freeenergyd01}(a)], and a point at a higher 
temperature [Fig.~\ref{fig:freeenergyd01}(b)].
In both cases the free energy presents four minima 
associated with distinct pairs of solutions
$(m_{\tau},m_{\sigma})$. The point at $(kT/J)=0.010$ presents
$(m_{\tau},m_{\sigma})=\{(-1,1);(0,1); (0,-1);(1,-1)\}$, whereas the 
point at $(kT/J)=0.387$ presents 
$(m_{\tau},m_{\sigma})=\{(-0.868, 0.991); (-0.100,0.991); (0.100, -0.991); 
(0.868, -0.991)\}$. 
The lower-temperature point represents a coexistence of the two phases 
shown in the case $D=0$ [cf. Fig.~\ref{fig:tdh00}(b)], namely, the 
ordered ($|m_{\tau}|=1$) and partially-ordered ($m_{\tau}=0$) phases. 
However, the higher-temperature point typifies the phenomenon
discussed in Fig.~\ref{fig:d01}, where distinct solutions with 
$|m_{\tau}|>0$ coexist, leading to a jump in this 
order parameter as one crosses the critical frontier, 
like illustrated in Fig.~\ref{fig:d01}(c) for the point 
$[(h_{0}/J)=0.59,(kT/J)=0.499]$. Although the 
magnetization $m_{\tau}$ presents a very 
curious behavior in topology II [cf., e.g.,  
Figs.~\ref{fig:d01}(b)--(d)], 
$m_{\sigma}$ remains essentially  
unchanged by the presence of the first-order 
critical frontier of 
Fig.~\ref{fig:d01}(a), as shown also in   
Fig.~\ref{fig:freeenergyd01}. 

%%%%%%%%%%%%%%%%%%%%%%%%%%%%
\begin{figure}[htp]
\begin{center}
%\includegraphics[width=0.45\textwidth,angle=0]{figures/figure6a.eps}
%\hspace{0.2cm}
%\includegraphics[width=0.45\textwidth,angle=0]{figures/figure6b.eps}
%\includegraphics[width=0.45\textwidth,angle=0]{figures/figure6a.pdf}
%\hspace{0.2cm}
%\includegraphics[width=0.45\textwidth,angle=0]{figures/figure6b.pdf}
%\includegraphics[height=5.5cm]{figures/figure6a.pdf}
%\hspace{0.2cm}
%\includegraphics[height=5.5cm]{figures/figure6b.pdf}
\includegraphics[height=7cm,angle=-90]{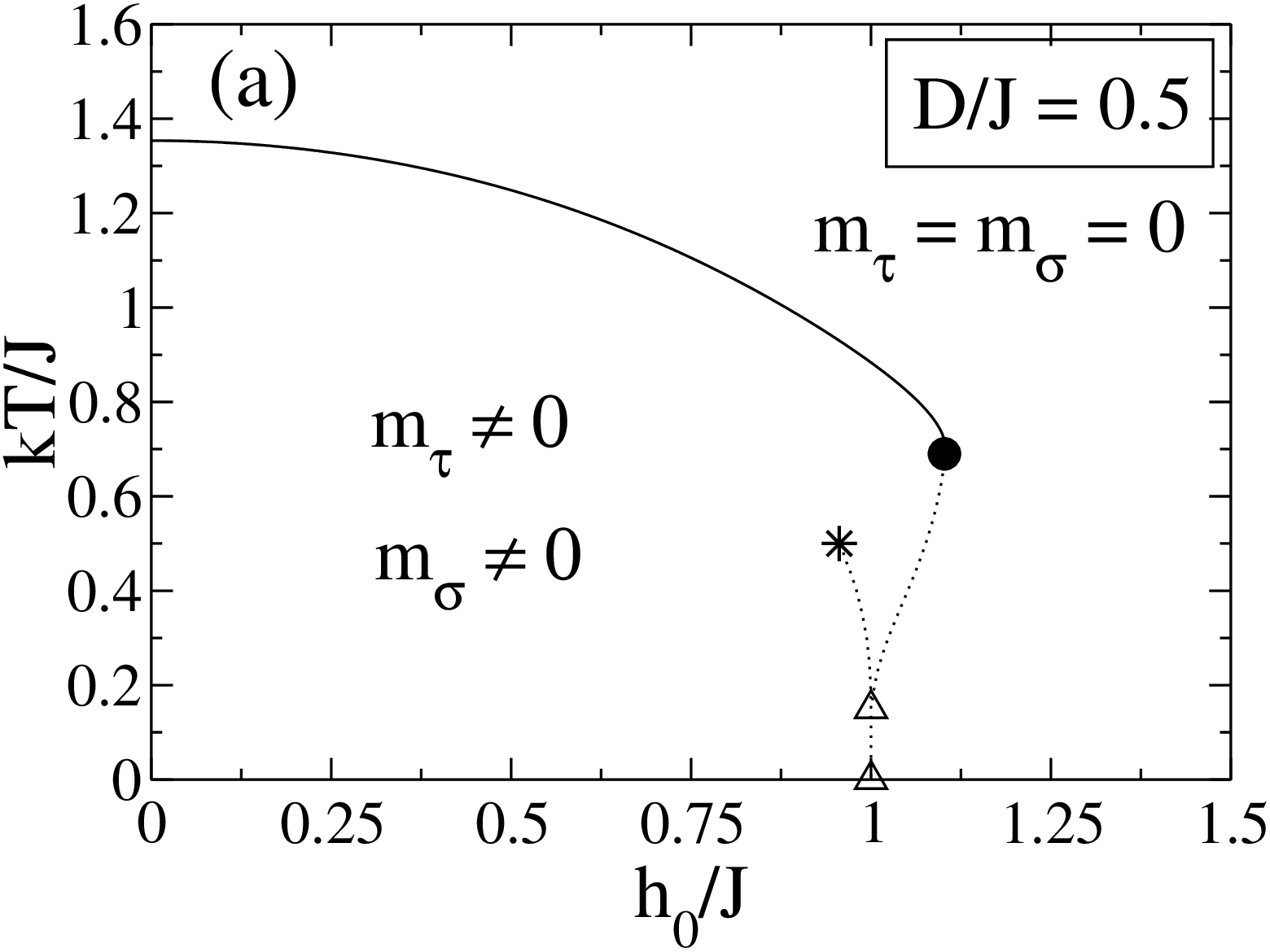}
\hspace{0.2cm}
\includegraphics[height=7cm,angle=-90]{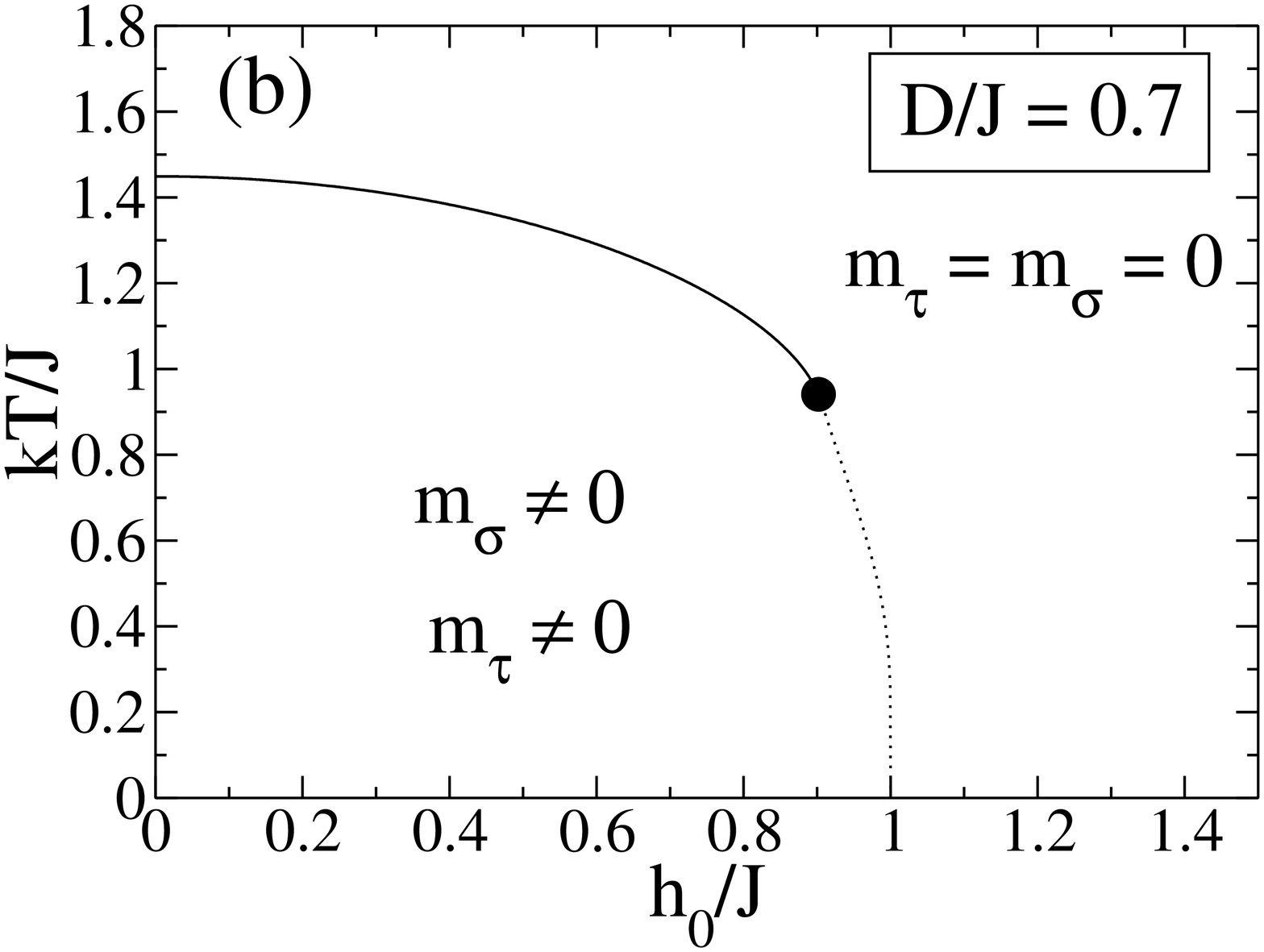}
\end{center}
\vspace{-.5cm}
\caption{Phase diagrams in the plane of dimensionless variables $kT/J$ 
versus $h_{0}/J$ for two different values of $D/J$:
(a) $(D/J)=0.5$, to be referred as topology III;
(b) $(D/J)=0.7$, to be referred as topology IV.} 
\label{fig:phasediagd0507}
\end{figure}
%%%%%%%%%%%%%%%%%%%%%%%%%%%%

In Fig.~\ref{fig:phasediagd0507} we present two other possible phase
diagrams, namely, the cases $(D/J)=0.5$ [Fig.~\ref{fig:phasediagd0507}(a), 
called herein topology III] and 
$(D/J)=0.7$ [Fig.~\ref{fig:phasediagd0507}(b), called herein topology IV]. 
Whereas topology III represents
a special situation that applies only for $(D/J)=0.5$, exhibiting the 
richest critical behavior of the present model, topology IV holds 
for any $(D/J)>0.5$.
In Fig.~\ref{fig:phasediagd0507}(a) one observes the appearance of
several multicritical points, denoted by the black circle (tricritical 
point), black asterisk (ordered critical point), and
empty triangles (triple points):
(i) The tricritical point, which signals the
encounter of the higher-temperature continuous phase transition
with the lower-temperature first-order phase transition,
found in the $D=0$ phase diagram [cf. Fig.~\ref{fig:tdh00}(b)],
have curiously disappeared for $0<(D/J)<0.5$, 
and emerged again for $(D/J)=0.5$;  
(ii) The ordered critical point exists for any $0 < (D/J) \leq 0.5$ 
[as shown in Fig.~\ref{fig:d01}(a)]; 
(iii) Two triple points, one at a finite temperature, whereas the 
other one occurs at zero temperature. It should be mentioned 
that such a zero-temperature triple point corresponds 
precisely to the one of Fig.~\ref{fig:groundstate}, at 
$(D/J)=0.5$ and $(h_{0}/J)=1.0$. 
The value $(D/J)=0.5$ is very special and will be considered as 
a threshold for both multicritical behavior and correlations 
between the two systems. We have observed that for  
$(D/J) \gtrsim 0.5$, the critical points shown in
Fig.~\ref{fig:phasediagd0507}(a) disappear, except for the
tricritical point that survives for 
$(D/J)>0.5$ [as shown in Fig.~\ref{fig:phasediagd0507}(b)].
Changes similar to those occurring 
herein between topologies II and III, as well as 
topologies III and IV, 
were found also in some 
magnetic systems, like the Ising and Blume-Capel
models, subject to random fields and/or 
dilution~\cite{kaufman,salmon1,salmon2,benyoussef,
carneiro,kaufmankanner}. 
Particularly, the splitting of the
low-temperature first-order critical frontier into 
two higher-temperature first-order lines that terminate
in the ordered and tricritical points, 
respectively [as exhibited in Fig.~\ref{fig:phasediagd0507}(a)],
is consistent with results found in  
the Blume-Capel model under  
a bimodal random magnetic, by 
varying the intensity of the crystal 
field~\cite{kaufmankanner}.

Another important feature of topology III concerns the 
lack of any type of 
magnetic order at finite temperatures for $(h_{0}/J)>1.1$, 
in contrast to the phase diagrams for 
$0 \leq (D/J) < 0.5$, for which there is $m_{\sigma} \neq 0$ 
for all $h_{0}/J$
[see, e.g., Figs.~\ref{fig:tdh00}(b) and~\ref{fig:d01}(a)]. 
This effect shows that $(D/J)=0.5$ represents a threshold value 
for the coupling between the variables $\{\sigma_{i}\}$ and 
$\{\tau_{i}\}$, so that for $(D/J) \geq 0.5$ the 
correlations among these variables become significant. 
As a consequence of these correlations, the fact 
of no magnetic 
order on the $\tau$-system ($m_{\tau} =0$) 
drives the the magnetization of the 
$\sigma$-system to zero as well, for $(h_{0}/J)>1.1$.    
It is important to notice that the $T=0$ phase diagram 
of Fig.~\ref{fig:groundstate}
presents a first-order critical line for $(D/J)=0.5$ and 
$(h_{0}/J)>1.0$, at which 
$m_{\tau} =0$, whereas in the $\sigma$-system both
$m_{\sigma}=0$ and  $|m_{\sigma}|=1$ minimize the Hamiltonian. 
By analyzing numerically the free-energy density 
of~\eq{eq:freeenergy} at low temperatures and $(h_{0}/J)>1.0$, 
we have verified that for any infinitesimal value of  
$kT/J$ destroys such a coexistence of solutions, leading to 
a minimum free energy at 
$m_{\tau}=m_{\sigma}=0 \ (\forall \, T>0)$. Consequently,
one finds that the low-temperature region in the interval  
$1.0 \leq (h_{0}/J) \leq 1.1$ becomes part of the {\bf P} phase.  
Hence, the phase diagram in 
Fig.~\ref{fig:phasediagd0507}(a) presents  
a reentrance phenomena for  
$1.0 \leq (h_{0}/J) \leq 1.1$. In this region, by lowering
the temperature gradually, one goes from a {\bf P} phase
to the ordered phase 
($m_{\tau} \neq 0$ ; $m_{\sigma} \neq 0$), and then back 
to the {\bf P} phase. This effect appears frequently 
in both theoretical and experimental investigations of 
disordered magnets~\cite{dotsenkobook,nishimoribook}. 

%%%%%%%%%%%%%%%%%%%%%%%%%%%%
\begin{figure}[htp]
\begin{center}
%\includegraphics[width=0.35\textwidth,angle=-90]{figures/figure7a.eps}
%\hspace{0.5cm} \vspace{0.7cm}
%\includegraphics[width=0.35\textwidth,angle=-90]{figures/figure7b.eps} \\
%\hspace{-2cm} \vspace{0.5cm}
%\includegraphics[width=0.4\textwidth,angle=0]{figures/figure7c.eps}
%\hspace{0.5cm} \vspace{0.5cm}
%\includegraphics[width=0.30\textwidth,angle=-90]{figures/figure7d.eps}
%\includegraphics[width=0.4\textwidth,angle=0]{figures/figure7a.pdf}
%\hspace{0.5cm} \vspace{0.7cm}
%\includegraphics[width=0.4\textwidth,angle=0]{figures/figure7b.pdf}
%\includegraphics[width=0.4\textwidth,angle=0]{figures/figure7c.pdf}
%\hspace{0.5cm}
%\includegraphics[width=0.4\textwidth,angle=0]{figures/figure7d.pdf}
%\includegraphics[height=5.2cm]{figures/figure7a.pdf}
%\hspace{0.5cm} \vspace{0.7cm}
%\includegraphics[height=5.2cm]{figures/figure7b.pdf}
%\includegraphics[height=5.2cm]{figures/figure7c.pdf}
%\hspace{0.5cm}
%\includegraphics[height=5.2cm]{figures/figure7d.pdf}
\includegraphics[height=7cm,clip,angle=-90]{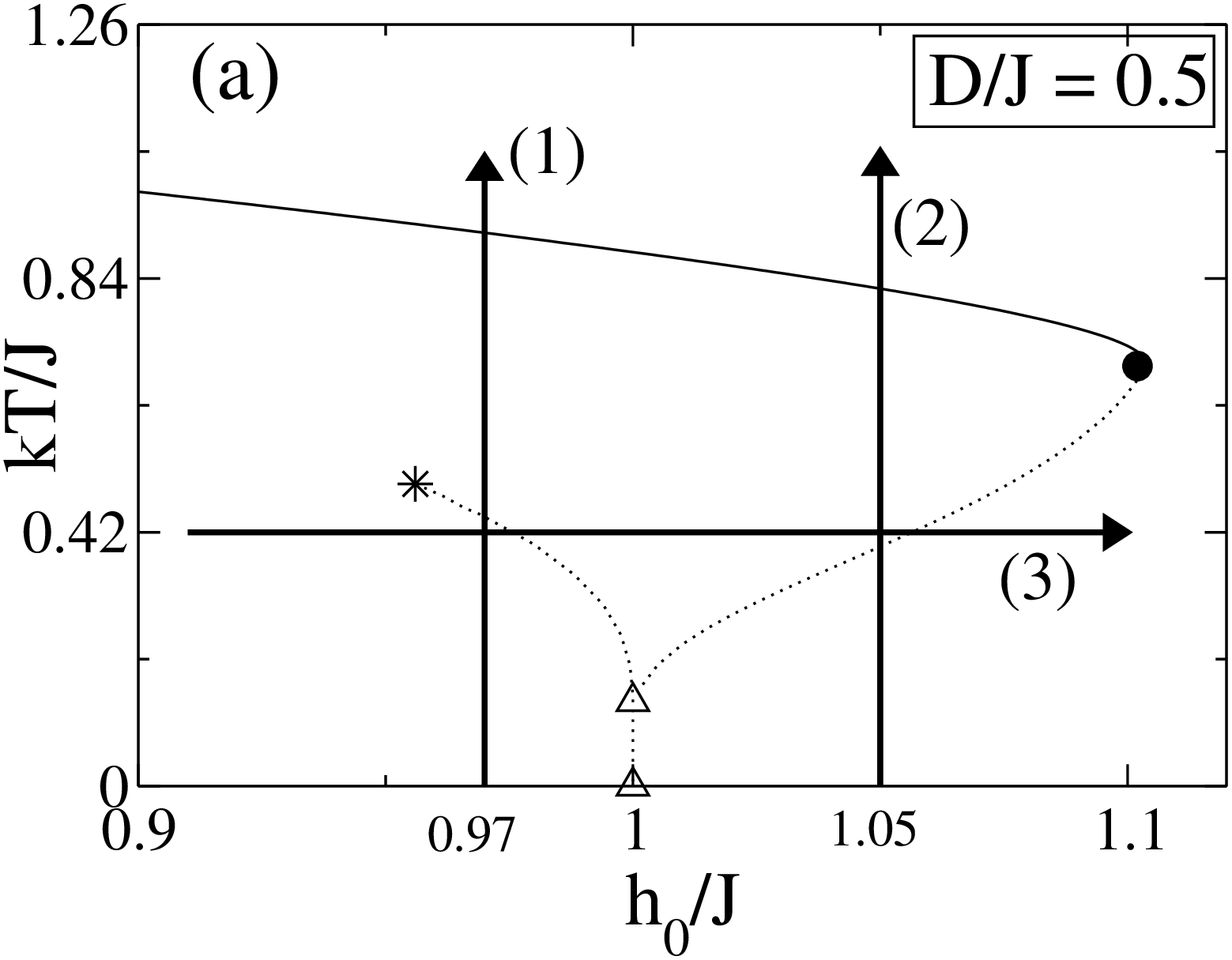}
\hspace{0.5cm} \vspace{0.7cm}
\includegraphics[height=7cm,clip,angle=-90]{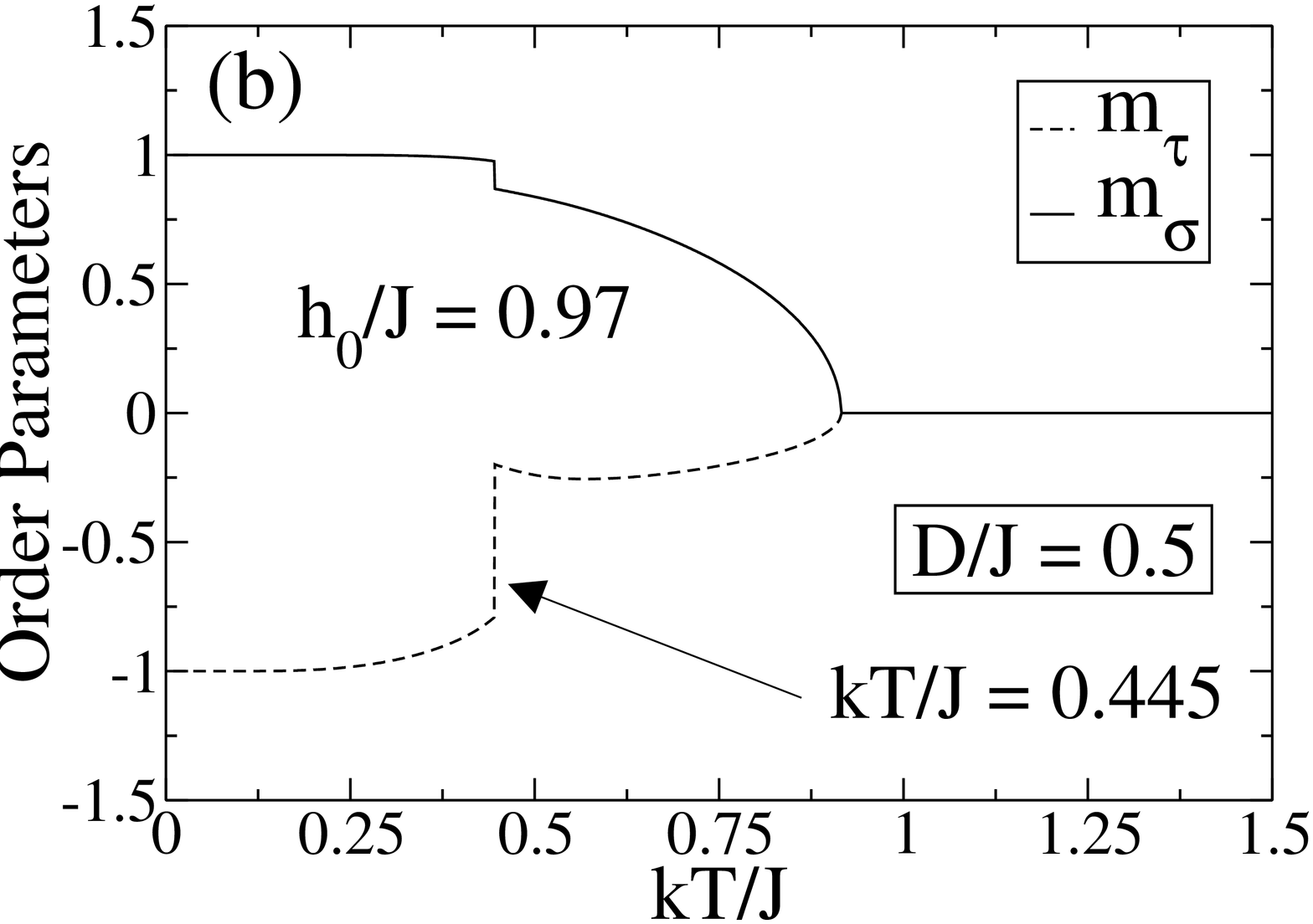} \\
\vspace{0cm} \hspace{-0.8cm}
\includegraphics[height=4.5cm,clip]{figure7c.eps}
\hspace{1.2cm}  
\includegraphics[height=4.5cm,clip]{figure7d.eps}
\end{center}
\vspace{0.2cm}
\caption{(a) The region of multicritical points of the phase diagram for 
$(D/J)=0.5$ [Fig.~\ref{fig:phasediagd0507}(a)] is amplified and three
thermodynamic paths are chosen for analyzing the magnetizations 
$m_{\tau}$ and $m_{\sigma}$.   
(b) Order parameters along thermodynamic path (1):  
$(h_{0}/J)=0.97$ and increasing temperatures.
(c) Order parameters along thermodynamic path (2):  
$(h_{0}/J)=1.05$ and increasing temperatures. 
(d) Order parameters along thermodynamic path (3):  
$(kT/J)=0.42$ and varying the field 
strength in the interval $0.9 \leq (h_{0}/J) \leq 1.15$.
Equivalent solutions exist by inverting the signs of 
$m_{\tau}$ and $m_{\sigma}$.}
\label{fig:magpaths123}
\end{figure}
%%%%%%%%%%%%%%%%%%%%%%%%%%%%

In Fig.~\ref{fig:magpaths123} we analyze the behavior of the  
$m_{\tau}$ and $m_{\sigma}$ for topology III, 
in the region of multicritical 
points of the phase diagram for 
$(D/J)=0.5$, along three typical thermodynamic paths, as 
shown in Fig.~\ref{fig:magpaths123}(a). 
In Fig.~\ref{fig:magpaths123}(b) we exhibit the behavior of
$m_{\tau}$ and $m_{\sigma}$ along path (1), where one 
sees that both parameters go through a jump by 
crossing the first-order critical line [$(kT/J)=0.445$], 
expressing a coexistence of different types 
of solutions for $m_{\tau}$ and $m_{\sigma}$ at this 
point. One notices a larger jump in $m_{\tau}$, so that 
to the right of the ordered critical point 
one finds a behavior similar to the one verified in topology II, 
where $|m_{\tau}|$ becomes very small, whereas  
$m_{\sigma}$ still presents significant values. 
Then, by further increasing the temperature, these parameters
tend smoothly to zero at the continuous critical frontier
separating the ordered and {\bf P} phases.    
In Fig.~\ref{fig:magpaths123}(c) we show the magnetizations
$m_{\tau}$ and $m_{\sigma}$ along path (2), 
within the region of the phase diagram 
where the reentrance phenomenon occurs; along this path, 
one increases the temperature, going from the {\bf P} phase 
to the ordered phase and then to the {\bf P} phase again.
Both parameters are zero for low enough temperatures,   
jumping to nonzero values at $(kT/J)=0.396$, as one 
crosses the first-order critical line. After such jumps,  
by increasing the temperature, these parameters
tend smoothly to zero at the border of the 
{\bf P} phase. The behavior shown in 
Fig.~\ref{fig:magpaths123}(c) confirm the reentrance 
effect, discussed previously.  
Finally, in Fig.~\ref{fig:magpaths123}(d) we exhibit 
the order parameters along thermodynamic path (3), 
for which the temperature is fixed at $(kT/J)=0.42$, with
the field varying in the range 
$0.9 \leq (h_{0}/J) \leq 1.15$. One sees that both 
magnetizations $m_{\tau}$ and $m_{\sigma}$ display 
jumps as one crosses each of the two first-order lines, 
evidencing a 
coexistence of different ordered states at the lower-temperature
jump, as well as a coexistence of the ordered and {\bf P} states
at the higher-temperature jump. 

The behavior presented by the order parameters in 
Figs.~\ref{fig:magpaths123}(b)--(d) shows clearly 
the fact that $(D/J)=0.5$ represents a threshold value 
for the coupling between the variables $\{\sigma_{i}\}$ and 
$\{\tau_{i}\}$. In all these cases, one sees that jumps
in the magnetization $m_{\sigma}$ are correlated with 
corresponding jumps in $m_{\tau}$.   
These results should be contrasted with those for the 
cases $(D/J)<0.5$, as illustrated  
in Fig.~\ref{fig:d01}(c), where a discontinuity 
in $m_{\tau}$ does not affect the smooth behavior presented
by $m_{\sigma}$.   

%%%%%%%%%%%%%%%%%%%%%%%%%%%%
\begin{figure}[htp]
\begin{center}
%\includegraphics[width=0.45\textwidth,angle=0]{figures/figure8a.eps}
%\hspace{0.2cm}
%\includegraphics[width=0.45\textwidth,angle=0]{figures/figure8b.eps}
%\includegraphics[width=0.45\textwidth,angle=0]{figures/figure8a.pdf}
%\hspace{0.2cm}
%\includegraphics[width=0.45\textwidth,angle=0]{figures/figure8b.pdf}
%\includegraphics[height=5.2cm]{figures/figure8d80a.pdf}
%\hspace{0.2cm}
%\includegraphics[height=5.2cm]{figures/figure8d80b.pdf}
\includegraphics[height=7cm,angle=-90]{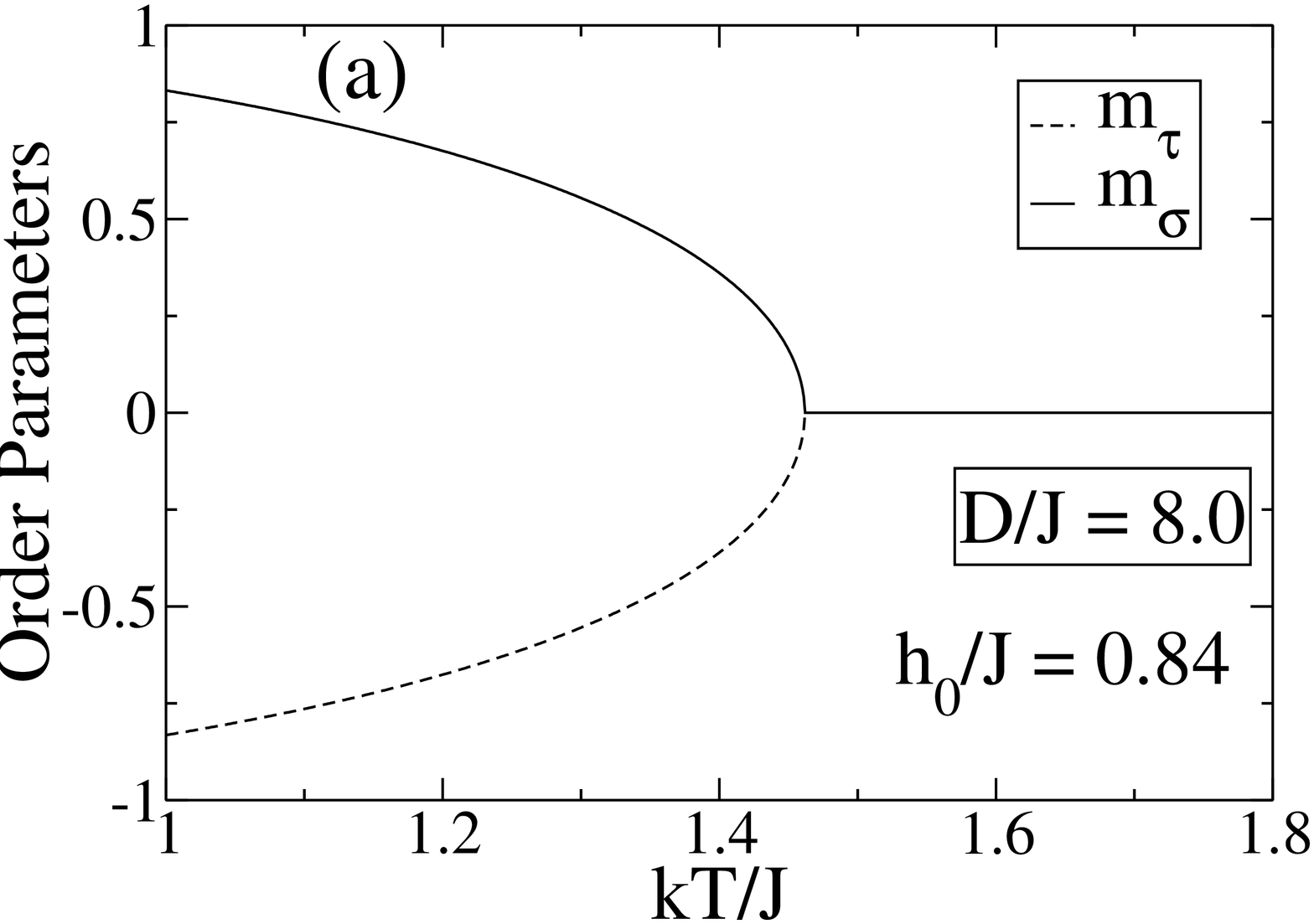}
\hspace{0.2cm}
\includegraphics[height=7cm,angle=-90]{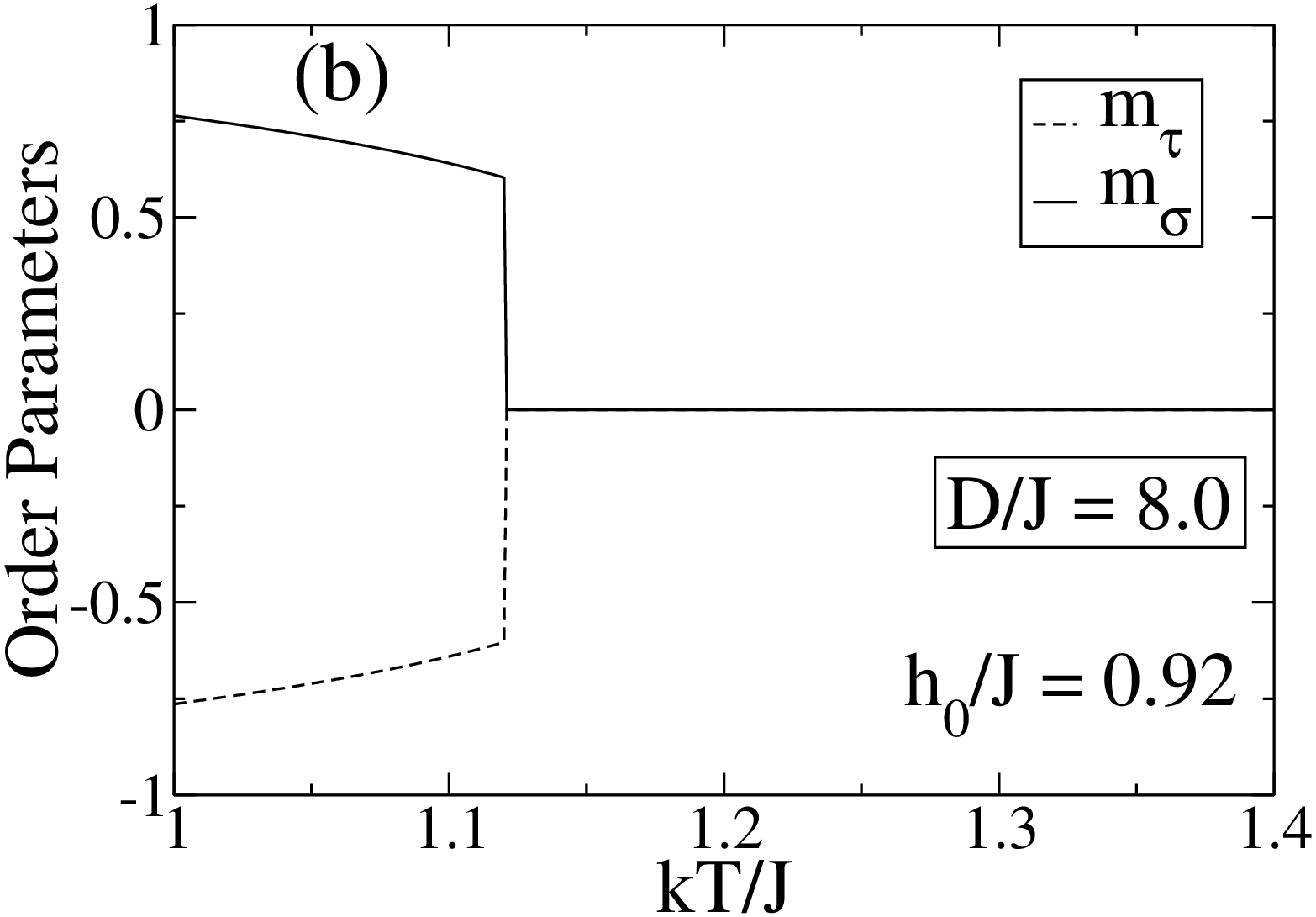}
\end{center}
\vspace{-.5cm}
\caption{The order parameters $m_{\tau}$ and $m_{\sigma}$
are represented versus the dimensionless temperature 
$kT/J$ for $(D/J)=8.0$ and two typical values of $h_{0}/J$: 
(a) Slightly to the left of the tricritical point; 
(b) Slightly to the right of the tricritical point. 
The associated phase diagram corresponds 
to topology IV [cf. Fig.~\ref{fig:phasediagd0507}(b)].
Equivalent solutions exist by inverting the signs of 
$m_{\tau}$ and $m_{\sigma}$.} 
\label{fig:magd80}
\end{figure}
%%%%%%%%%%%%%%%%%%%%%%%%%%%%

The phase diagram shown in Fig.~\ref{fig:phasediagd0507}(b), 
which corresponds to topology IV, is valid for any
for any $(D/J)>0.5$. Particularly, 
the critical point where the low-temperature 
first-order critical 
frontier touches the zero-temperature axis 
is kept at $(h_{0}/J)=1$, for all $(D/J)>0.5$, 
in agreement with Fig.~\ref{fig:groundstate}. 
We have verified  
only quantitative changes in such a phase diagram 
by increasing 
the coupling between the variables $\{\sigma_{i}\}$ and 
$\{\tau_{i}\}$. Essentially, the whole continuous critical 
frontier moves towards 
higher temperatures, leading to an increase in the values 
of the critical temperature 
for $(h_{0}/J)=0$, as well as in the temperature 
associated with the tricritical point, whereas the abscissa 
of this point remains typically unchanged. Moreover,
in what concerns the order parameters, 
the difference between $|m_{\tau}|$ 
and $m_{\sigma}$ decreases, in such a way 
that for $(D/J) \rightarrow \infty$, one obtains 
$m_{\tau}=-m_{\sigma}$. 
This later effect is illustrated in 
Fig.~\ref{fig:magd80}, where we represent the 
order parameters $m_{\tau}$ and $m_{\sigma}$
versus temperature, for a sufficiently large value of
$D/J$, namely, $(D/J)=8.0$, in   
two typical choices of $h_{0}/J$, close 
to the tricritical point. 
In Fig.~\ref{fig:magd80}(a) $m_{\tau}$ and $m_{\sigma}$
are analyzed slightly to the left of the tricritical 
point, exhibiting the usual continuous behavior, 
whereas in Fig.~\ref{fig:magd80}(b) they 
are considered slightly to the right of the tricritical
point, presenting jumps as one crosses the
first-order critical frontier. However, 
the most important conclusion from Fig.~\ref{fig:magd80} 
concerns the fact that in both cases one has essentially  
$m_{\tau}=-m_{\sigma}$, showing that  the random 
field applied solely to the $\tau$-system influences the 
$\sigma$-system in a similar way, due to the 
high value of $D/J$ considered.  
We have verified that for $(D/J)=8.0$
the two systems become so strongly 
correlated, such that 
$m_{\tau}=-m_{\sigma}$ holds along 
the whole phase diagram, 
within our numerical accuracy. 

\section{Conclusions}

We have analyzed the effects of a coupling $D$ 
between two Ising models, defined in terms of variables
$\{\tau_{i}\}$ and $\{\sigma_{i}\}$.  
The model was considered in the limit of infinite-range
interactions, where all spins in each system
interact by means of an exchange coupling $J>0$, typical 
of ferromagnetic interactions.
Motivated by a qualitative description of
systems like plastic crystals, 
the variables $\{\tau_{i}\}$ and $\{\sigma_{i}\}$ would 
represent rotational and translational degrees 
of freedom, respectively. Since the rotational 
degrees of freedom are expected to change more 
freely than the translational ones,  
a random field acting only on the variables
$\{\tau_{i}\}$ was considered.
For this purpose, a bimodal random field, 
$h_{i} = \pm h_{0}$, with equal probabilities,
was defined on the $\tau$-system.
The model was investigated through its free energy 
and its two order parameters, namely, 
$m_{\tau}$ and $m_{\sigma}$. 
 
We have shown that such a system presents a very rich 
critical behavior, depending on the particular choices 
of $D/J$ and $h_{0}/J$. 
Particularly, at zero temperature, the phase diagram in the plane 
$h_{0}/J$ versus $D/J$ exhibits ordered, partially-ordered, 
and disordered phases. This phase diagram is symmetric 
around $(D/J)=0$, so that for sufficiently low values of      
$h_{0}/J$ one finds ordered phases characterized by 
$m_{\sigma}=m_{\tau}=\pm 1$ ($D<0$) and 
$m_{\sigma}=-m_{\tau}=\pm 1$ ($D>0$). 
We have verified that $|D/J|=1/2$
plays an important role in the present model, such
that at zero temperature one has the disordered 
phase ($m_{\sigma}=m_{\tau}=0$)
for  $|D/J|>1/2$ and $(h_{0}/J)>1$. 
Moreover, the partially-ordered phase, 
where $m_{\sigma}=\pm 1$ and $m_{\tau}=0$, 
occurs for $(h_{0}/J)>1/2+|D/J|$ and $|D/J|<1/2$.  
In this phase diagram all phase transitions are 
of the first-order type, and three triple points were found. 
In the case of plastic crystals, 
the sequence of transitions from the disordered 
to the partially-ordered, and then to the
ordered phases, would correspond to the 
sequence of transitions from the liquid to
the plastic crystal, and then to ordered crystal
phases.

Due to the symmetry around $D=0$, the 
finite-temperature phase diagrams were considered
only for $D>0$, for which the ordered
phase was identified by $m_{\sigma}>0$ and
$m_{\tau}<0$, whereas the partially-ordered phase
by $m_{\sigma}>0$ and
$m_{\tau}=0$ (equivalent solutions also exist by 
inverting the signs of these order parameters).   
Several phase diagrams in the 
plane $kT/J$ versus $h_{0}/J$ were studied, 
by varying gradually $D/J$. We have found
four qualitatively different types of phase diagrams, 
denominated as topologies I [$(D/J)=0$], II [$0<(D/J)<1/2$], 
III [$(D/J)=1/2$], and IV [$(D/J)>1/2$]. Such a
classification reflects the fact that $(D/J)=1/2$ 
represents a threshold value 
for the coupling between the variables $\{\sigma_{i}\}$ and 
$\{\tau_{i}\}$, so that for $(D/J) \geq 1/2$ the 
correlations among these variables become significant, 
as verified through the behavior of the order parameters
$m_{\tau}$ and $m_{\sigma}$. 
From all these cases, only topology IV 
typifies a well-known phase diagram, 
characterized by a tricritical point, where the 
higher-temperature continuous frontier meets 
the lower-temperature first-order critical line.
This phase diagram is qualitatively similar to 
the one found for the  
Ising ferromagnet in the presence of a bimodal 
random field~\cite{aharony}, and it does not 
present the herein physically relevant 
partially-ordered phase. 
For $(D/J) \geq 1/2$, even though the random field
is applied only in the $\tau$-system, the correlations 
lead the $\sigma$-system to follow a qualitatively
similar behavior. 
  
The phase diagrams referred as topologies I and II 
exhibit all three phases. In the later case we have found 
a first-order critical line terminating at an ordered
critical point, leading to the potential physical realization
of going continuously from the ordered phase to the 
partially-ordered phase by circumventing this 
critical point.
In these two topologies, the sequence of transitions 
from the disordered 
to the partially-ordered, and then to the
ordered phase, represents the physical 
situation that occurs in plastic crystals. 
For conveniently chosen thermodynamic paths, 
i.e., varying temperature and random field appropriately,
one may go from the liquid phase 
($m_{\sigma}=m_{\tau}=0$), to a plastic-crystal phase
($m_{\sigma} \neq 0$; $m_{\tau}=0$), where the rotational degrees 
of freedom are found in a disordered state, and then, 
to an ordered crystal phase
($m_{\sigma} \neq 0$; $m_{\tau} \neq 0$).
 
 From the point of view of multicritical behavior, 
topology III [$(D/J)=1/2$] corresponds to 
the richest type of phase diagram, being 
characterized by several critical lines and 
multicritical points; one finds its most  
complex criticality around $(h_{0}/J)=1$, signaling 
a great competition among the different types of orderings.  
Although the partially-ordered phase 
does not appear in this particular case, one has also 
the possibility of circumventing the ordered critical point,  
such as to reach a region of the phase diagram 
along which $|m_{\tau}|$ becomes very small,
resembling a partially-ordered phase.  

Since the infinite-range interactions among 
variables of each Ising system correspond to a limit
where mean-field approach becomes exact, an immediate  
question concerns whether some of the results obtained above
represent an artifact of such limit. 
Certainly, such a relevant point is directly related with the 
existence of some of these features in the associated
short-range three-dimensional magnetic models. For example, the
tricritical point found in topologies III and IV is essentially 
the same that appears within the mean-field approach of the 
Ising model in the presence of a bimodal random field. 
This later model has been extensively investigated on a cubic 
lattice through different numerical approaches, where the 
existence of this tricritical point is still very controversial. 
On the other hand, a first-order critical frontier terminating
at an ordered critical point, and the fact that one can
go from one phase to another by
circumventing this point, represents a typical 
physical situation that occurs in real
substances. The potential for exhibiting such a 
relevant feature represents an important advantage 
of the present model. 
 
Finally, we emphasize that the rich critical behavior presented 
in the phase diagrams corresponding to topologies II and III
suggest the range $0<(D/J) \leq 1/2$ as appropriate
for describing plastic crystals. 
The potential of exhibiting successive transitions from the 
ordered to the partially-ordered and then to the 
disordered phase should be useful for a better 
understanding of these systems. 
Furthermore, the characteristic   
of going continuously from the ordered phase 
to the partially-ordered phase by circumventing an ordered 
critical point represents a typical physical situation that 
occurs in many substances, 
and opens the possibility for 
the present model to describe a wider range of materials. 

\vskip 2\baselineskip

{\large\bf Acknowledgments}

\vskip \baselineskip
\noindent
 The partial financial supports from CNPq, 
FAPEAM-Projeto-Universal-Amazonas, 
and FAPERJ (Brazilian agencies) are acknowledged. 

\vskip 2\baselineskip

\end{document}